\documentclass{llncs}
\usepackage{makeidx}  
\usepackage{amsmath,amsfonts,amssymb}
\usepackage{graphicx}
\usepackage[misc]{ifsym}
\usepackage[tight]{subfigure}
\usepackage{caption} 
\newcommand{\vA}{\boldsymbol{A}}
\newcommand{\vf}{\boldsymbol{f}}

\newcommand{\vp}{\boldsymbol{p}}
\newcommand{\vP}{\boldsymbol{P}}
\newcommand{\wTV}{||\boldsymbol{f}||_{\text{wTV}}}
\newcommand{\subjt}{\text{ subject to }}

\begin{document}
\frontmatter          
\pagestyle{headings}  
\mainmatter              
\title{Data Consistent Artifact Reduction for Limited Angle Tomography with Deep Learning Prior}
\titlerunning{Data Consistent Artifact Reduction with Deep Learning}  
%
\author{Yixing~Huang\inst{1}{(\Letter)} \and
Alexander Preuhs\inst{1} \and G\"unter Lauritsch\inst{2} \and Michael Manhart\inst{2} \and Xiaolin Huang\inst{3}
\and Andreas~Maier\inst{1,4}
}

\authorrunning{Yixing Huang et al.} 

%
%
\institute{Pattern Recognition Lab, Friedrich-Alexander-Universität Erlangen-Nürnberg, 91058 Erlangen, Germany \\
\email{yixing.yh.huang@fau.de} \\
\and
Siemens Healthcare GmbH, 91301 Forchheim, Germany\\
\and
Institute of Image Processing and Pattern Recognition, Shanghai Jiao Tong University, 200240 Shanghai, China\\
\and
Erlangen Graduate School in Advanced Optical Technologies
(SAOT), 91058 Erlangen, Germany}

%
%
\maketitle              

\begin{abstract}
Robustness of deep learning methods for limited angle tomography is challenged by two major factors: 
a) due to insufficient training data the network may not generalize well to unseen data; 
b) deep learning methods are sensitive to noise.
Thus, generating reconstructed images directly from a neural network appears inadequate. We propose to constrain the reconstructed images to be consistent with the measured projection data, while the unmeasured information is complemented by learning based methods. For this purpose, a data consistent artifact reduction (DCAR) method is introduced: 
First, a prior image is generated from an initial limited angle reconstruction via deep learning as a substitute for missing information. 
Afterwards, a conventional iterative reconstruction algorithm is applied, integrating the data consistency in the measured angular range and the prior information in the missing angular range. 
This ensures data integrity in the measured area, while inaccuracies incorporated by the deep learning prior lie only in areas where no information is acquired. 
The proposed DCAR method achieves significant image quality improvement:  for $120^\circ$ cone-beam limited angle tomography more than $10\%$ RMSE reduction in noise-free case and more than $24\%$ RMSE reduction in noisy case compared with a state-of-the-art U-Net based method.

\end{abstract}
\section{Introduction}
Recently, deep learning has achieved overwhelming success in various computed tomography (CT) applications \cite{andreas2019gentle,wang2018image}, including low-dose CT \cite{kang2017deep,chen2017low,yang2018low}, sparse-view reconstruction \cite{han2016deep,han2018framing,chen2018learn}, and metal artifact reduction \cite{gjesteby2017deep,zhang2018convolutional}. 
In this work, we are interested in the application of deep learning to limited angle tomography. Image reconstruction from data acquired in an insufficient angular range is called limited angle tomography. It arises when the gantry rotation of a CT system is restricted by other system parts, or a super short scan is preferred for the sake of quick scanning time, low dose, or less contrast agent. 

Conventionally, limited angle tomography is addressed by extrapolation methods \cite{louis1980picture,huang2017restoration} or iterative reconstruction algorithms with total variation \cite{sidky2008image,chen2013limited,huang2018scale}. In the past three years, various deep learning methods have been investigated in limited angle tomography \cite{wurfl2016deep,zhang2016image,gu2017multi,huang2018some,wurfl2018deep,Bubba2019Learning}.   
For example, Gu and Ye adapted the U-Net architecture \cite{ronneberger2015u} to learn artifacts from streaky images in the multi-scale wavelet domain \cite{gu2017multi}. Good quality images are obtained by this method for $120^\circ$ limited angle tomography. The results presented in the literature reveal promising developments for a clinical applicability of deep learning-based reconstructions.  

However, the robustness of deep learning in practical applications is still a concern. On one hand, deep learning methods may fail to generalize to new test instances as these methods are trained only on an insufficient dataset. On the other hand, due to the curse of high dimensional space \cite{goodfellow2014explaining}, deep neural networks have been reported to be vulnerable to small perturbations, including adversarial examples and noise \cite{szegedy2013intriguing,yuan2017adversarial,antun2019instabilities}. In the field of limited angle tomography, our previous work \cite{huang2018some} has demonstrated that the U-Net method is not robust to Poisson noise as well. In this work, we devise an algorithm overcoming these limitations by enforcing data consistency with the measured raw data.

Since generating reconstructed images directly from a neural network appears inadequate, we propose to combine deep learning with known operators. 
The first category of such approaches is to build deep neural network architectures directly based on analytic formulas of conventional methods. In these neural networks, each layer represents a certain known operator whose weights are fine tuned by data-driven learning to improve precision. Therefore, they are called ``precision learning" \cite{Christopher2018precision,maier2019learning}. In precision learning, maximal error bounds are limited by prior information of the analytic formulas. W\"urfl et al. \cite{wurfl2016deep,wurfl2018deep} proposed a neural network architecture based on filtered back-projection (FBP) to learn the compensation weights \cite{riess2013tv} for limited angle reconstruction. However, this particular method is not suitable for small angular ranges, e.\,g. $120^\circ$ cone-beam limited angle tomography, since no redundant data are available to compensate missing data.
 The second category is to use deep learning and conventional methods to reconstruct different parts of an imaged object respectively. 
Bubba et al. \cite{Bubba2019Learning} proposed a hybrid deep learning-shearlet framework for limited angle tomography, where an iterative shearlet transform algorithm \cite{frikel2013sparse} is utilized to reconstruct visible singularities of an imaged object while a U-Net based neural network with dense blocks \cite{huang2017densely} is utilized to predict the invisible ones. This method achieves better image quality than pure model or data-driven-based reconstruction methods.
 The third category is to use deep learning results as prior information for conventional methods. Zhang et al's method \cite{zhang2018convolutional} is such an example for metal artifact reduction. To make the best of measured data, Zhang et al. used deep learning predictions as prior images to interpolate projection data in metal corrupted areas \cite{zhang2018convolutional}.

In this work, we choose the third category for limited angle tomography. 
In \cite{huang2018some}, the U-Net learns artifacts from streaky images in the image domain only. Reconstruction images obtained by such image-to-image prediction are very likely not consitent to measured data as the prediction does not have any direct connection to measured data. To make predicted images data consistent, a data consistent artifact reduction (DCAR) method is proposed: The predicted images are used as prior images to provide information in missing angular ranges first; Afterwards, a conventional reconstruction algorithm is applied to integrate the prior information in the missing angular ranges and constrain the reconstruction images to be consistent to the measured data in the acquired angular range.

\section{Method}
\subsection{The U-Net architecture}
\begin{figure}[tbh]
\centering
\includegraphics[width = 0.9\textwidth]{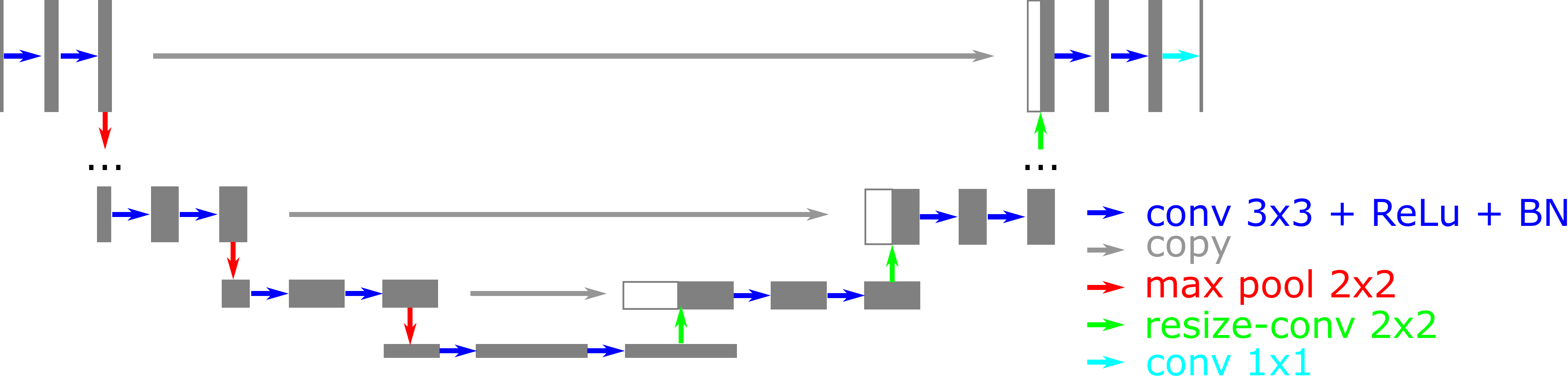}
\caption{The U-Net architecture for limited angle tomography (modified from \cite{ronneberger2015u}).}
\label{Fig:UNetArch}
\end{figure}

As displayed in Fig.~\ref{Fig:UNetArch}, the same U-Net architecture as that in \cite{huang2018some} is used for artifact reduction in limited angle tomography, which is modified from \cite{ronneberger2015u} and \cite{gu2017multi}. 
In this work, the input images are Ram-Lak-kernel-based FBP reconstructions from limited angle data, while the output images are artifact images. The Hounsfield scaled images are normalized to ensure stable training. An $\ell_2$ loss function is used.

\subsection{Data consistent artifact reduction}

\textbf{Data fidelity of measured data:}
We denote measured projections by $\vp_{\text{m}}$ and the system matrix for the measured projections by $\vA_{\text{m}}$ in cone-beam limited angle tomography. The FBP reconstruction from the measured data $\vp_{\text{m}}$ only is denoted by $\vf_{\text{FBP}}$. The artifact image, predicted by the U-Net, is denoted by $\vf_{\text{artifact}}$. Then an estimation of the artifact-free image, denoted by $\vf_{\text{U-Net}}$, is obtained by $\vf_{\text{U-Net}} = \vf_{\text{limited}} - \vf_{\text{artifact}}$. Due to insuficient training data or sensitivity to noise in the application of limited angle tomography \cite{huang2018some}, $\vf_{\text{U-Net}}$ is not consistent to the measured data. A data consistent reconstruction image $\vf$ follows the following constraint,
\begin{equation}
||\vA_{\text{m}}\vf - \vp_{\text{m}}|| < e_1,
\label{eqn:dataConsistencye1}
\end{equation}
where $e_1$ is a parameter for error tolerance. When the measured data $\vp_{\text{m}}$ are noise-free, $e_1$ is ideally zero. When $\vp_{\text{m}}$ contains noise caused by various physical effects, $e_1$ is a certain positive value. 

Because of the severe ill-posedness of limited angle tomography, the number of images satisfying the above constraint is not unique. We aim to reconstruct an image which satisfies the above constraint and meanwhile is close the U-Net reconstruction $\vf_{\text{U-Net}}$. For this purpose, 
we choose to initialize the image $\vf$ with $\vf_{\text{U-Net}}$ and solve it in an iterative manner, i.\,e.,
\begin{equation}
||\vA_{\text{m}}\vf - \vp_{\text{m}}|| < e_1, \text{ and }\vf^{(0)} = \vf_{\text{U-Net}}.
\label{eqn:dataConsistencye2}
\end{equation}
In this way, the data consistency constraint is fully satisfied. Note that with such initialization, the deep learning prior $\vf_{\text{U-Net}}$ contributes to the selection of one image among all images satisfying Eqn.~(\ref{eqn:dataConsistencye1}).

\textbf{Data fidelity of unmeasured data:}
 We further denote the system matrix for an unmeasured angular range by $\vA_\text{u}$ and its corresponding projections by $\vp_{\text{u}}$. In cone-beam computed tomography, a short scan is necessary for image reconstruction. Therefore, in this work, we choose $\vA_\text{u}$ such that $\vA_{\text{m}}$ and $\vA_\text{u}$ form a system matrix for a short scan CT system. Although the projections $\vp_{\text{u}}$ are not measured, they can be approximated by the deep learning reconstruction $\vf_{\text{U-Net}}$ via forward projection. Making the best of such prior information, the following constraint is proposed,
 \begin{equation}
 ||\vA_{\text{u}}\vf - \vA_{\text{u}} \vf_{\text{U-Net}}||=|| \vA_{\text{u}}(\vf - \vf_{\text{U-Net}})|| < e_2,
 \label{eqn:deepLearningConsistencye2}
\end{equation}  
 where the error tolerance parameter $e_2$ accounts for the inaccuracy of the deep learning prior $\vf_{\text{U-Net}}$. When $\vf_{\text{U-Net}}$ has bad image quality, a relative large value should be set. This constraint indicates that the final reconstruction $\vf$ is close to the deep learning prior $\vf_{\text{U-Net}}$ in the unmeasured space and the difference between them is controlled by the parameter $e_2$. 
 
\textbf{Regularization:} 
To further reduce noise and artifacts corresponding to the error tolerance of $e_1$ and $e_2$, additional regularization is applied. In this work, the following iterative reweighted total variation (wTV) regularization \cite{huang2018scale} is utilized,
\begin{equation}
\begin{split}
&||\boldsymbol{f}^{(n)}||_{\text{wTV}}=\sum_{x,y,z}\boldsymbol{w}^{(n)}_{x,y,z}||\mathcal{D}\boldsymbol{f}^{(n)}_{x,y,z}||,\\
&\boldsymbol{w}^{(n)}_{x,y,z}=\frac{1}{||\mathcal{D}\boldsymbol{f}^{(n-1)}_{x,y,z}||+\epsilon},
\end{split}
\label{eq:WeightsUpdate}
\end{equation}
where $\boldsymbol{f}^{(n)}$ is the image at the $n^\text{th}$ iteration, $\boldsymbol{w}^{(n)}$ is the weight vector for the $n^\text{th}$ iteration which is computed from the previous iteration, and $\epsilon$ is a small positive value added to avoid division by zero. A smaller value of $\epsilon$ results in finer image resolution but slower convergence speed.

\textbf{Overall algorithm:}
Therefore, the overall objective function for our DCAR method is as the following,
 \begin{equation}
 \min\wTV, \subjt \left\lbrace
 \begin{array}{l}
 \vf^{(0)} = \vf_{\text{U-Net}},\\
 ||\vA_{\text{m}}\vf - \vp_{\text{m}}|| < e_1,\\
 ||\vA_{\text{u}}\vf - \vA_{\text{u}} \vf_{\text{U-Net}}|| < e_2,
 \end{array} 
 \right.
 \label{eqn:ObjectiveDataConsistentDeepLearning}
 \end{equation}
 which is a constrained optimization problem. 
 
 To solve the above objective function, simultaneous algebraic reconstruction technique (SART) + wTV is applied \cite{huang2018scale}, i.\,e., SART is utilized to minimize the data fidelity terms of Eqns.~(\ref{eqn:dataConsistencye1}) and (\ref{eqn:deepLearningConsistencye2}), while a gradient descent method is utilized to minimize the wTV term. To minimize the data fidelity terms, SART is adapted as the following,
\begin{equation}
\vf_j^{(l+1)}=
\left\lbrace
\begin{array}{ll}
\vf_j^{(l)}+\lambda \cdot \frac{\sum_{\vp_i\in \vP_{\beta}}\frac{\mathcal{S}_{e_1}\left(\vp_i-\sum^N_{k=1}\vA_{i,k}\cdot \vf_k^{(l)}\right)}{\sum_{k=1}^{N}\vA_{i,k}}\cdot \vA_{i,j}}{\sum_{\vp_i\in \vP_{\beta}}\vA_{i,j}},&\text{ if }\vp_i \text{ is measured},\\
\vf_j^{(l)}+\lambda \cdot \frac{\sum_{\vp_i\in \vP_{\beta}}\frac{\mathcal{S}_{e_2}\left(\sum^N_{k=1}\vA_{i,k}\cdot \left(\vf_{\text{U-Net}}-\vf_k^{(l)}\right)\right)}{\sum_{k=1}^{N}\vA_{i,k}}\cdot \vA_{i,j}}{\sum_{\vp_i\in \vP_{\beta}}\vA_{i,j}},&\text{ otherwise},
\end{array}
\right.
\label{eqn:SART2}
\end{equation}
 where the system matrix $\vA$ is the combination of $\vA_{\text{m}}$ and $\vA_{\text{u}}$, the projection vector $\vp$ is the combination of $\vp_{\text{m}}$ and $\vp_{\text{u}}$, and $\mathcal{S}_{\tau}$ is a soft-thresholding operator with threshold $\tau$ to deal with error tolerance. $\vp_{\text{u}}$ is estimated and substituted by $\vA_{\text{u}} \vf_{\text{U-Net}}$ in the above formula. For other parameters, $\vf_j$ stands for the $j^\text{th}$ pixel of $\vf$, $\vp_i$ stands for the $i^\text{th}$ projection ray of $\vp$, $\vA_{i,j}$ is the element of $\vA$ at the $i^\text{th}$ row and the $j^\text{th}$ column, $l$ is the iteration number, $N$ is the total pixel number of $\vf$, $\lambda$ is a relaxation parameter, $\beta$ is the X-ray source rotation angle, and $\vP_{\beta}$ stands for the set of projection rays when the source is at rotation angle $\beta$. To minimize the wTV term, the gradient of $\wTV$ w.\,r.\,t.~each pixel is computed and a gradient descent method using backtracking line search is applied \cite{huang2018scale}.

\subsection{Experimental setup}
We validate the proposed DCAR algorithm using 17 patients' data from the AAPM Low-Dose CT Grand Challenge \cite{mccollough2017low} simulated in $120^\circ$ cone-beam limited angle tomography without and with Poisson noise.

\textbf{System configuration:}
For each patient's data, limited angle projections are simulated in a cone-beam limited angle tomography system with parameters listed in Table~\ref{tab:ConeBeamParametersLowDoseCTData}. In the noisy case, Poisson noise is simulated considering an initial exposure of $10^5$ photons at each detector pixel before attenuation.

\begin{table}[t]
\begin{center}
\begin{tabular}{|l|c|}
\hline
Parameter & Value\\
\hline
Scan angular range & $120^\circ (210^\circ)$\\
\hline
Start angle & $30^\circ (0^\circ)$\\
\hline
End Angle & $150^\circ (210^\circ)$\\
\hline
Angular step & $1^\circ$\\
\hline
Source-to-detector distance & 1200.0\,mm\\
\hline
Source-to-isocenter distance & 600.0\,mm\\
\hline
Detector size & $620 \times 480$\\
\hline
Detector pixel size & 1.0\,mm $\times$ 1.0\,mm\\
\hline
Image size & $256 \times 256 \times 256$\\
\hline
Image pixel size & 1.25\,mm $\times$ 1.25\,mm $\times$ 1.0\,mm\\
\hline

\end{tabular}
\caption{The system configuration of cone-beam limited angle tomography to validate the proposed DCAR algorithm, where the angular parameters in the brackets are for a short scan configuration.}
\label{tab:ConeBeamParametersLowDoseCTData}
\end{center}
\vspace{-20pt}
\end{table}

\textbf{Training and test data:}
To investigate the dependence of the U-Net's performance on training data, leave-one-out cross validation is performed. For each validation, data from 16 patients are used for training while the data from the remaining one are used for test. Among the 16 patients, 25 slices from each patient are chosen for training. For the validation patient, all the 256 slices from the FBP reconstruction $\vf_{\text{FBP}}$ are fed to the U-Net for evaluation. As the artifacts are mainly caused by limited angle scan, the effect of cone-beam angle is neglected. Therefore, 2-D slices are used for training and test instead of volumes to avoid high computation. Both the training and test data are noise-free in the noise-free case, while both the training data and test data contain Poisson noise in the noisy case.
 
 \textbf{Algorithm parameters:}
 The U-Net is trained on the above data using the Adam optimizer. The learning rate is $10^{-3}$ for the first 100 epochs, $10^{-4}$ for the $101-130^{\text{th}}$ epochs, and $10^{-5}$ for the $131-150^\text{th}$ epochs. The $\ell_2$-norm is applied to regularize the network weights. The regularization parameter is $10^{-4}$.

For reconstruction, in the noise-free case, the error tolerance value $e_1$ is set to 0.001 in Eqn.~(\ref{eqn:dataConsistencye1}) for discretization error, while $e_1$ is set to 0.01 for the noisy case. The U-Net reconstructions $\vf_\text{U-Net}$ of each patient are reprojected in the angular range of $[0^\circ, 210^\circ]$. Other system parameters are the same as those in Table.~\ref{tab:ConeBeamParametersLowDoseCTData}. A relatively large tolerance value of 0.5 is chosen empirically for $e_2$ in Eqn.~(\ref{eqn:deepLearningConsistencye2}). For SART, the parameter $\lambda$ in Eqn.~(\ref{eqn:SART2}) is set to 0.8. For the wTV regularization, the parameter $\epsilon$ is set to 5\,HU for weight update. 50 iterations of SART + wTV are applied using the U-Net reconstruction $\vf_{\text{U-Net}}$ as initialization to get the final reconstruction. For comparison, the results of 100 iterations of SART + wTV using zero images as initialization are presented.

\section{Results}
\begin{figure}[tbh]
\centering
\begin{minipage}{0.19\linewidth}
\centering
$\vf_{\text{reference}}$
\end{minipage}
\begin{minipage}{0.19\linewidth}
\centering
$\vf_{\text{FBP}}$
\end{minipage}
\begin{minipage}{0.19\linewidth}
\centering
$\vf_{\text{wTV}}$
\end{minipage}
\begin{minipage}{0.19\linewidth}
\centering
$\vf_{\text{U-Net}}$
\end{minipage}
\begin{minipage}{0.19\linewidth}
\centering
$\vf_{\text{DCAR}}$
\end{minipage}

\vspace{3pt}

\begin{minipage}{0.19\linewidth}
\subfigure[ ]{
\includegraphics[width = \linewidth]{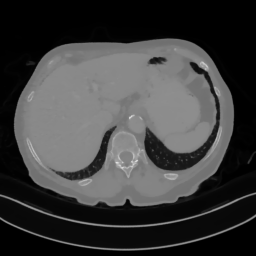}
\label{subfig:referenceP18S128}
}
\end{minipage}
\begin{minipage}{0.19\linewidth}
\subfigure[328\,HU]{
\includegraphics[width = \linewidth]{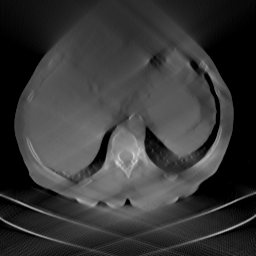}
\label{subfig:reconLimitedP18S128}
}
\end{minipage}
\begin{minipage}{0.19\linewidth}
\subfigure[ 138\,HU ]{
\includegraphics[width = \linewidth]{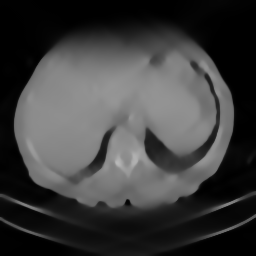}
\label{subfig:reconwTV18S128}
}
\end{minipage}
\begin{minipage}{0.19\linewidth}
\subfigure[ 105\,HU ]{
\includegraphics[width = \linewidth]{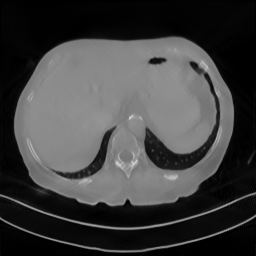}
\label{subfig:reconUNetP18S128}
}
\end{minipage}
\begin{minipage}{0.19\linewidth}
\subfigure[ 88\,HU ]{
\includegraphics[width = \linewidth]{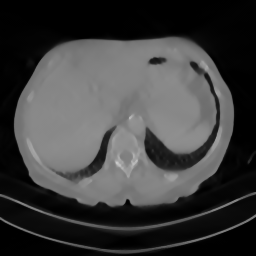}
\label{subfig:reconDCARP18S128}
}
\end{minipage}

\begin{minipage}{0.19\linewidth}
\subfigure[]{
\includegraphics[width = \linewidth]{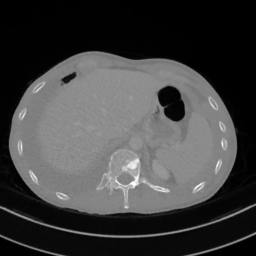}
\label{subfig:referenceP5S125}
}
\end{minipage}
\begin{minipage}{0.19\linewidth}
\subfigure[ 333\,HU]{
\includegraphics[width = \linewidth]{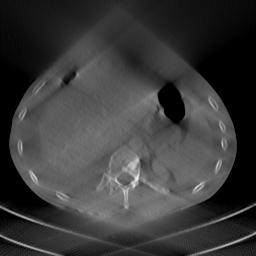}
\label{subfig:reconLimitedP5S125}
}
\end{minipage}
\begin{minipage}{0.19\linewidth}
\subfigure[ 134\,HU]{
\includegraphics[width = \linewidth]{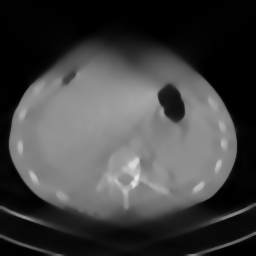}
\label{subfig:reconwTVP5S125}
}
\end{minipage}
\begin{minipage}{0.19\linewidth}
\subfigure[ 71\,HU ]{
\includegraphics[width = \linewidth]{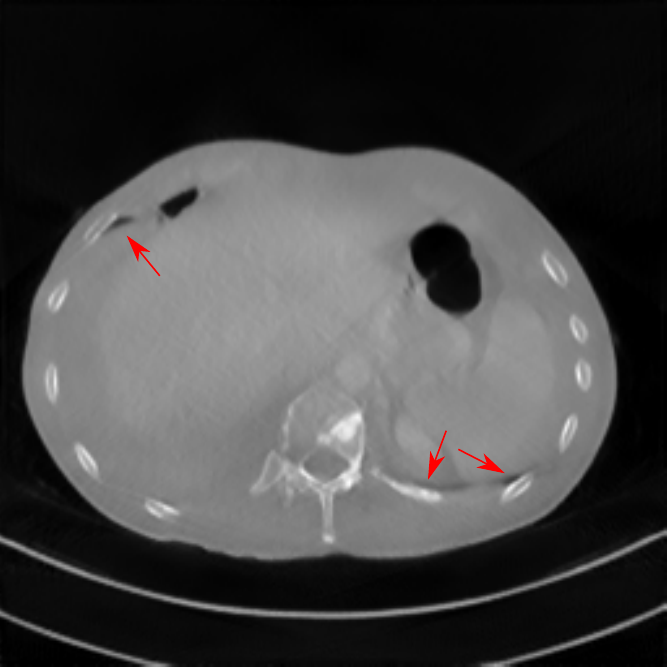}
\label{subfig:reconUNetP5S125}
}
\end{minipage}
\begin{minipage}{0.19\linewidth}
\subfigure[ 63\,HU ]{
\includegraphics[width = \linewidth]{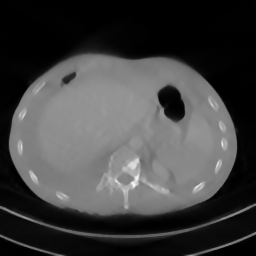}
\label{subfig:reconDCARP5S125}
}
\end{minipage}

\begin{minipage}{0.19\linewidth}
\subfigure[]{
\includegraphics[width = \linewidth]{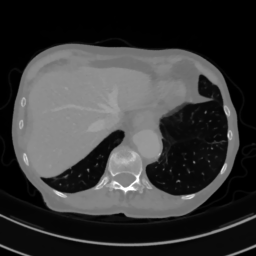}
 \label{subfig:referenceP8S70}
}
\end{minipage}
\begin{minipage}{0.19\linewidth}
\subfigure[ 317\,HU]{
\includegraphics[width = \linewidth]{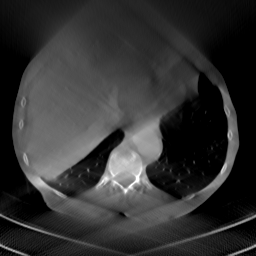}
 \label{subfig:reconLimitedP8S70}
}
\end{minipage}
\begin{minipage}{0.19\linewidth}
\subfigure[ 140\,HU]{
\includegraphics[width = \linewidth]{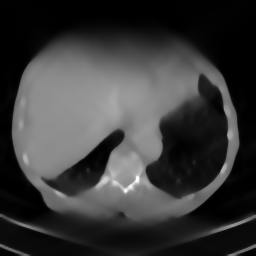}
\label{subfig:reconwTVP8S70}
}
\end{minipage}
\begin{minipage}{0.19\linewidth}
\subfigure[ 122\,HU]{
\includegraphics[width = \linewidth]{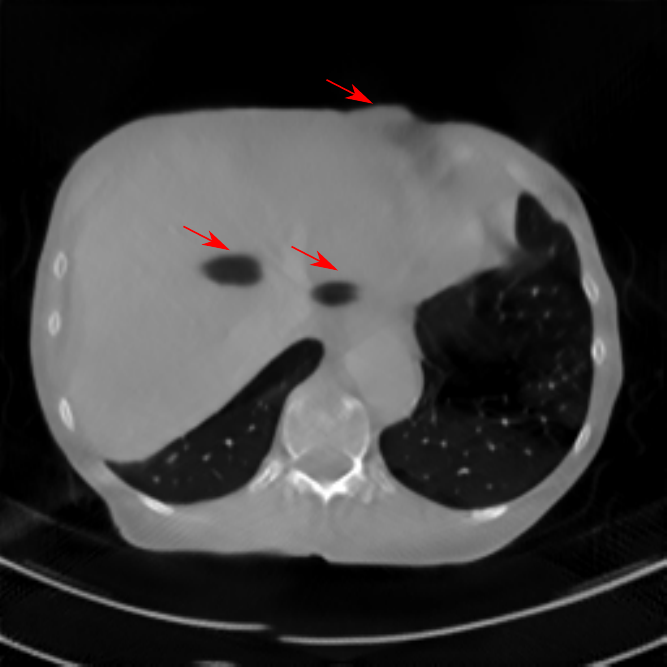}
 \label{subfig:reconUNetP8S70}
}
\end{minipage}
\begin{minipage}{0.19\linewidth}
\subfigure[ 85\,HU ]{
\includegraphics[width = \linewidth]{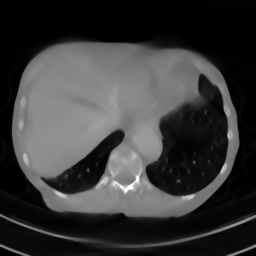}
\label{subfig:reconDCARP8S70}
}
\end{minipage}
\caption{Reconstruction results of three example slices by U-Net and DCAR in noise-free $120^\circ$ cone-beam limited angle tomography. The images from top to bottom are from Patient NO. 17, 4, and 7, respectively. The areas marked by the arrows are reconstructed incorrectly by the U-Net, which are rectified by DCAR. The RMSE value for each slice is displayed in their subtitle. Window: [-1000, 1000]\,HU.}
\label{Fig:DCARResults}
\end{figure}
The results of three example slices in $120^\circ$ noise-free cone-beam limited angle tomography are displayed in Fig.~\ref{Fig:DCARResults}. These three slices are from Patient NO. 17, 4, and 7, respectively. In each row, the reference image $\vf_{\text{reference}}$, the FBP reconstruction $\vf_{\text{FBP}}$, the U-Net reconstruction $\vf_{\text{U-Net}}$, the SART + wTV (using wTV for short in the following) reconstruction $\vf_{\text{wTV}}$, and the DCAR reconstruction $\vf_{\text{DCAR}}$ are displayed in order. Comparing Fig.~\ref{subfig:reconLimitedP18S128} with Fig.~\ref{subfig:referenceP18S128}, the body outline of Patient 17 is severely distorted due to missing data. Moreover, many streaks occur, obscuring anatomical structures such as the ribs and the vertebra. Figs.~\ref{subfig:reconwTV18S128}-\ref{subfig:reconDCARP18S128} demonstrate that wTV, U-Net, and DCAR all are able to improve these corrupted anatomical structures. The root-mean-square error (RMSE) is reduced significantly from 328\,HU for $\vf_{\text{FBP}}$ to 138\,HU for $\vf_{\text{wTV}}$ w.\,r.\,t.~the reference image. But the intensity values at the top body part are still too low in $\vf_{\text{wTV}}$. The RMSE is further reduced to 105\,HU for $\vf_{\text{U-Net}}$ in Fig.~\ref{subfig:reconUNetP18S128}, while DCAR reaches the smallest RMSE value of 88\,HU for this slice. In the middle row and the bottom row, the U-Net is able to reconstruct most anatomical structures well. However, the structures indicated by the red arrows are apparently incorrect compared with reference images. In Fig.~\ref{subfig:reconUNetP5S125}, the dark holes indicated by the red arrows appear, very likely because the corresponding areas in Fig.~\ref{subfig:reconLimitedP5S125} have low intensities due to dark streak artifacts. In contrast, the two large holes in Fig.\,\ref{subfig:reconUNetP8S70} occur without any clear clue, since no dark areas are present in Fig.\,\ref{subfig:reconLimitedP8S70}. Apparently Fig.~\ref{subfig:reconUNetP8S70} is not consistent to measured data and by using DCAR, these two holes are reestablished, although some darkness remains.

\begin{figure}[h!]
\centering
\begin{minipage}{1\linewidth}
\includegraphics[width = 1\linewidth]{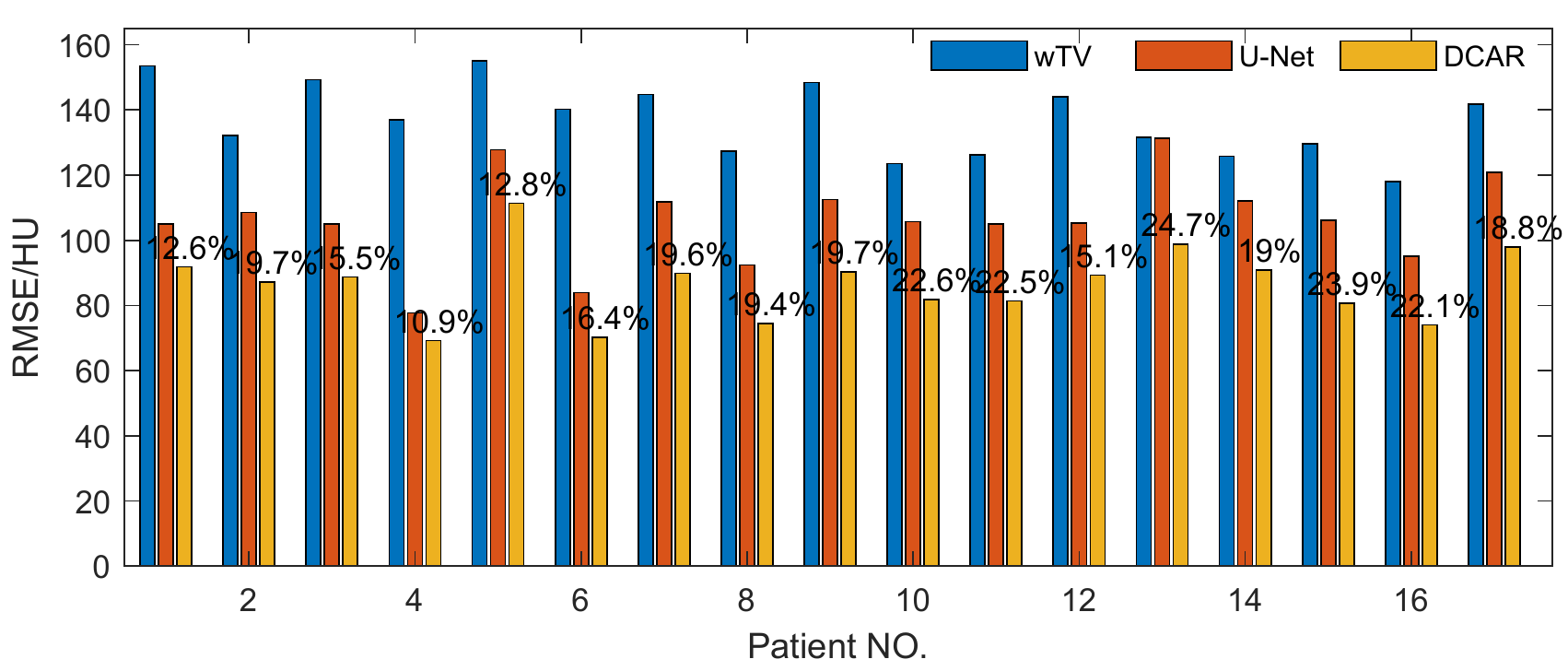}
\end{minipage}
\caption{Comparison of the mean RMSE values by wTV, U-Net, and DCAR for each patient in $120^\circ$ noise-free cone-beam limited angle tomography. The relative improvement of DACR from the U-Net is marked for each patient.
}
\label{Fig:barPlotMeanValues}
\end{figure}

The comparison of the mean RMSE values for wTV, U-Net, and DCAR in the leave-one-out cross-validation is plotted in Fig.~\ref{Fig:barPlotMeanValues}. It indicates that wTV has the largest mean RMSE values among these three methods and DCAR achieves more than $10\%$ improvement in mean RMSE values compared with the U-Net. This convincingly demonstrates the benefit of DCAR in reducing artifacts for limited angle tomography.

\begin{figure}[tbh]
\centering
\begin{minipage}{0.19\linewidth}
\centering
$\vf_{\text{reference}}$
\end{minipage}
\begin{minipage}{0.19\linewidth}
\centering
$\vf_{\text{FBP}}$
\end{minipage}
\begin{minipage}{0.19\linewidth}
\centering
$\vf_{\text{wTV}}$
\end{minipage}
\begin{minipage}{0.19\linewidth}
\centering
$\vf_{\text{U-Net}}$
\end{minipage}
\begin{minipage}{0.19\linewidth}
\centering
$\vf_{\text{DCAR}}$
\end{minipage}

\vspace{3pt}

\begin{minipage}{0.19\linewidth}
\subfigure[]{
\includegraphics[width = \linewidth]{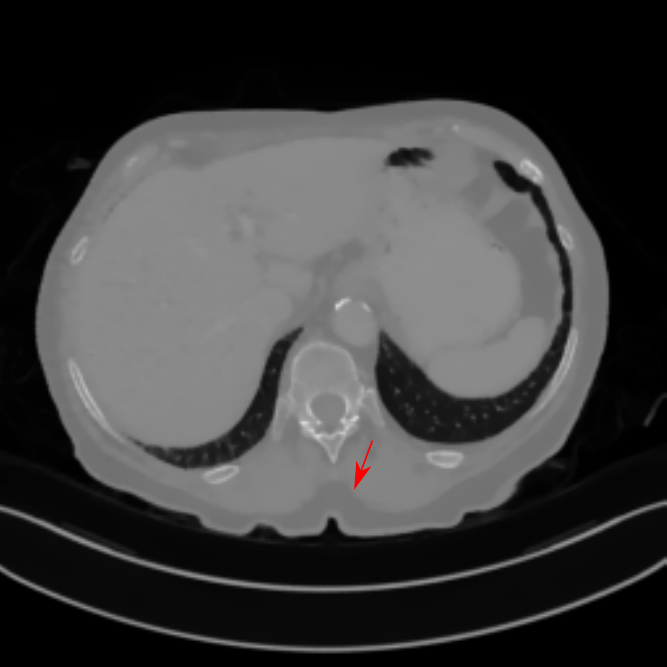}
 \label{subfig:referenceP18S1282}
}
\end{minipage}
\begin{minipage}{0.19\linewidth}
\subfigure[ 358\,HU]{
\includegraphics[width = \linewidth]{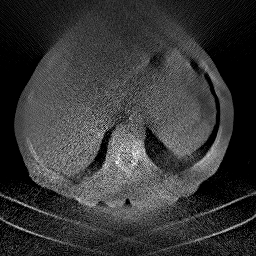}
 \label{subfig:reconLimitedP18S128Poi}
}
\end{minipage}
\begin{minipage}{0.19\linewidth}
\subfigure[ 141\,HU]{
\includegraphics[width = \linewidth]{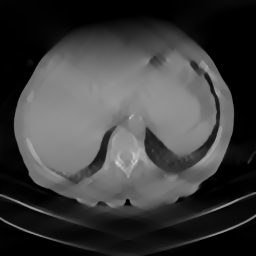}
 \label{subfig:reconwTVP18S128Poi}
}
\end{minipage}
\begin{minipage}{0.19\linewidth}
\subfigure[ 138\,HU]{
\includegraphics[width = \linewidth]{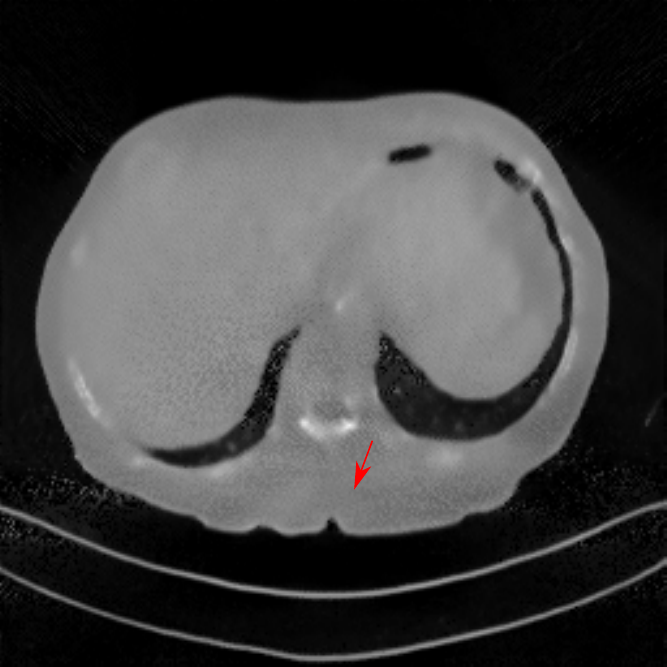}
 \label{subfig:reconUNetP18S128Poi}
}
\end{minipage}
\begin{minipage}{0.19\linewidth}
\subfigure[102\,HU ]{
\includegraphics[width = \linewidth]{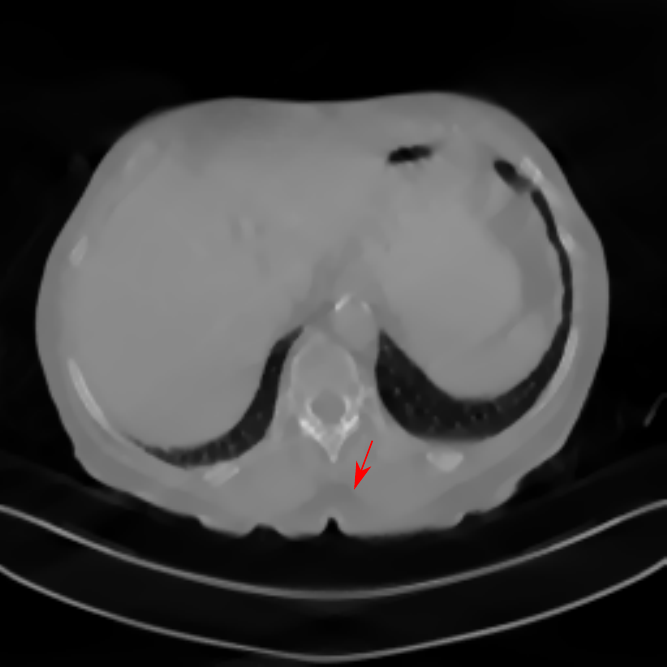}
\label{subfig:reconDCARP18S128Poi}
}
\end{minipage}

\begin{minipage}{0.19\linewidth}
\subfigure[ ]{
\includegraphics[width = \linewidth]{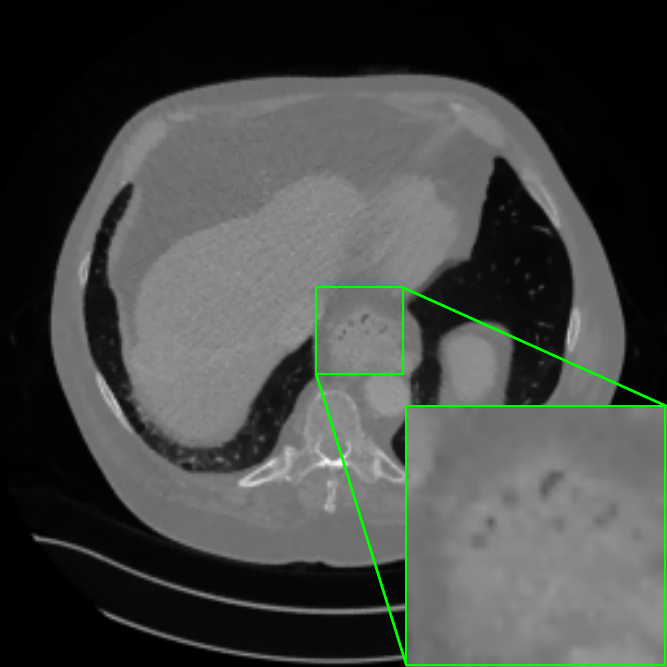}
\label{subfig:referenceP2S94Poi}
}
\end{minipage}
\begin{minipage}{0.19\linewidth}
\subfigure[ 299\,HU]{
\includegraphics[width = \linewidth]{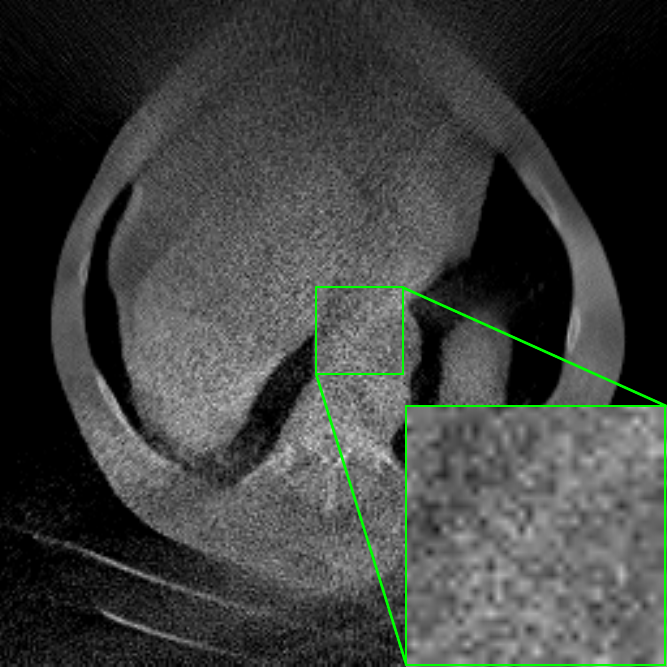}
\label{subfig:reconLimitedP2S94Poi}
}
\end{minipage}
\begin{minipage}{0.19\linewidth}
\subfigure[ 123\,HU]{
\includegraphics[width = \linewidth]{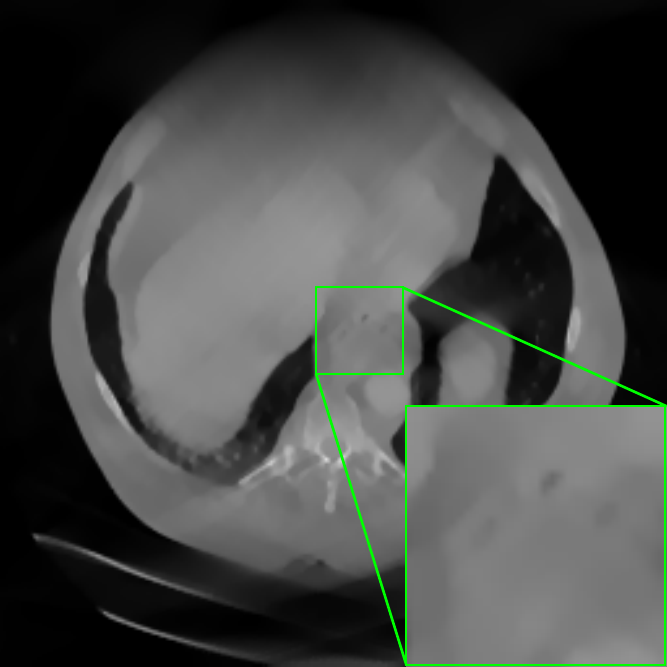}
\label{subfig:reconwTVP2S94Poi}
}
\end{minipage}
\begin{minipage}{0.19\linewidth}
\subfigure[114\,HU ]{
\includegraphics[width = \linewidth]{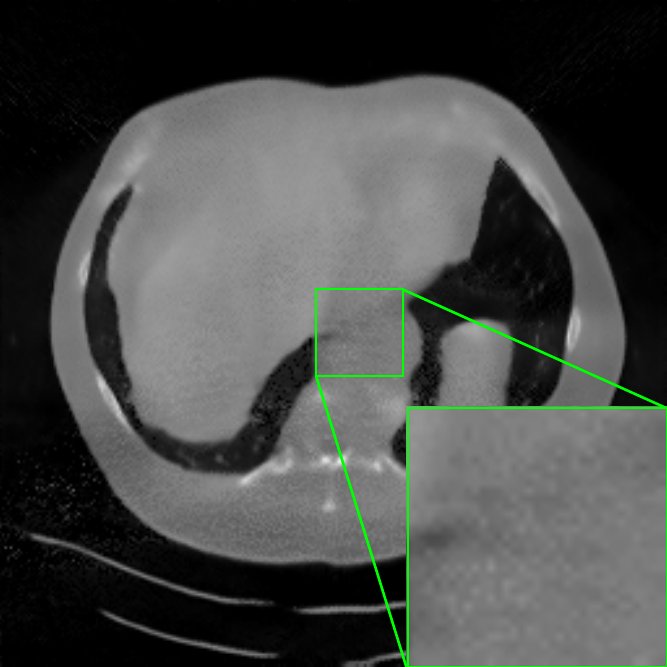}
\label{subfig:reconUNetP2S94Poi}
}
\end{minipage}
\begin{minipage}{0.19\linewidth}
\subfigure[85\,HU ]{
\includegraphics[width = \linewidth]{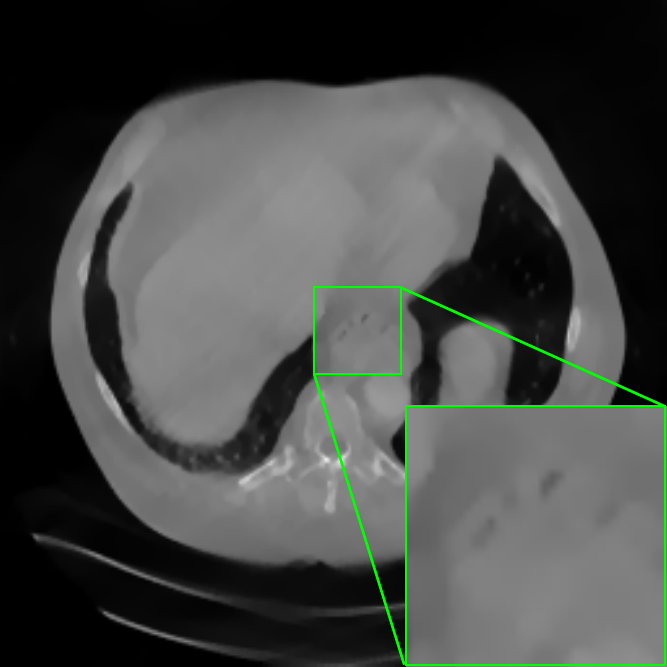}
\label{subfig:reconDCARP2S94Poi}
}
\end{minipage}

\begin{minipage}{0.19\linewidth}
\subfigure[ ]{
\includegraphics[width = \linewidth]{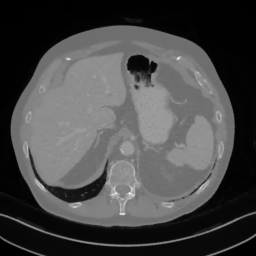}
\label{subfig:referenceP9S140Poi}
}
\end{minipage}
\begin{minipage}{0.19\linewidth}
\subfigure[ 379\,HU]{
\includegraphics[width = \linewidth]{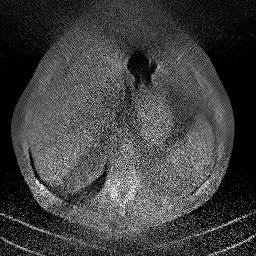}
\label{subfig:reconLimitedP9S140Poi}
}
\end{minipage}
\begin{minipage}{0.19\linewidth}
\subfigure[ 120\,HU]{
\includegraphics[width = \linewidth]{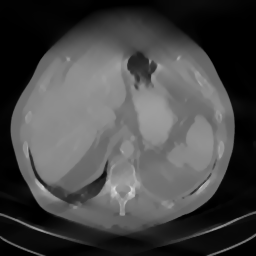}
\label{subfig:reconwTVP9S140Poi}
}
\end{minipage}
\begin{minipage}{0.19\linewidth}
\subfigure[125\,HU ]{
\includegraphics[width = \linewidth]{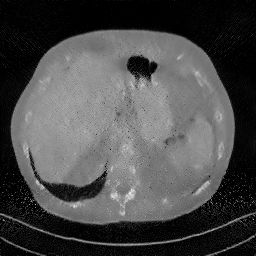}
\label{subfig:reconUNetP9S140Poi}
}
\end{minipage}
\begin{minipage}{0.19\linewidth}
\subfigure[74\,HU ]{
\includegraphics[width = \linewidth]{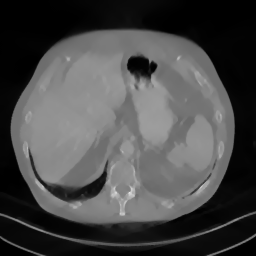}
\label{subfig:reconDCARP9S140Poi}
}
\end{minipage}
\caption{Reconstruction results of three example slices by U-Net and DCAR in $120^\circ$ cone-beam limited angle tomography with Poisson noise. The images from top to bottom are from Patient NO. 17, 2, and 8, respectively. The areas marked by the arrows are reconstructed incorrectly by the U-Net, which are rectified by DCAR. The RMSE value for each slice is displayed in their subtitle. Window: [-1000, 1000]\,HU.}
\label{Fig:DCARResultsPoisson}
\end{figure}

\begin{figure}[h!]
\centering
\begin{minipage}{1\linewidth}
\includegraphics[width = 1\linewidth]{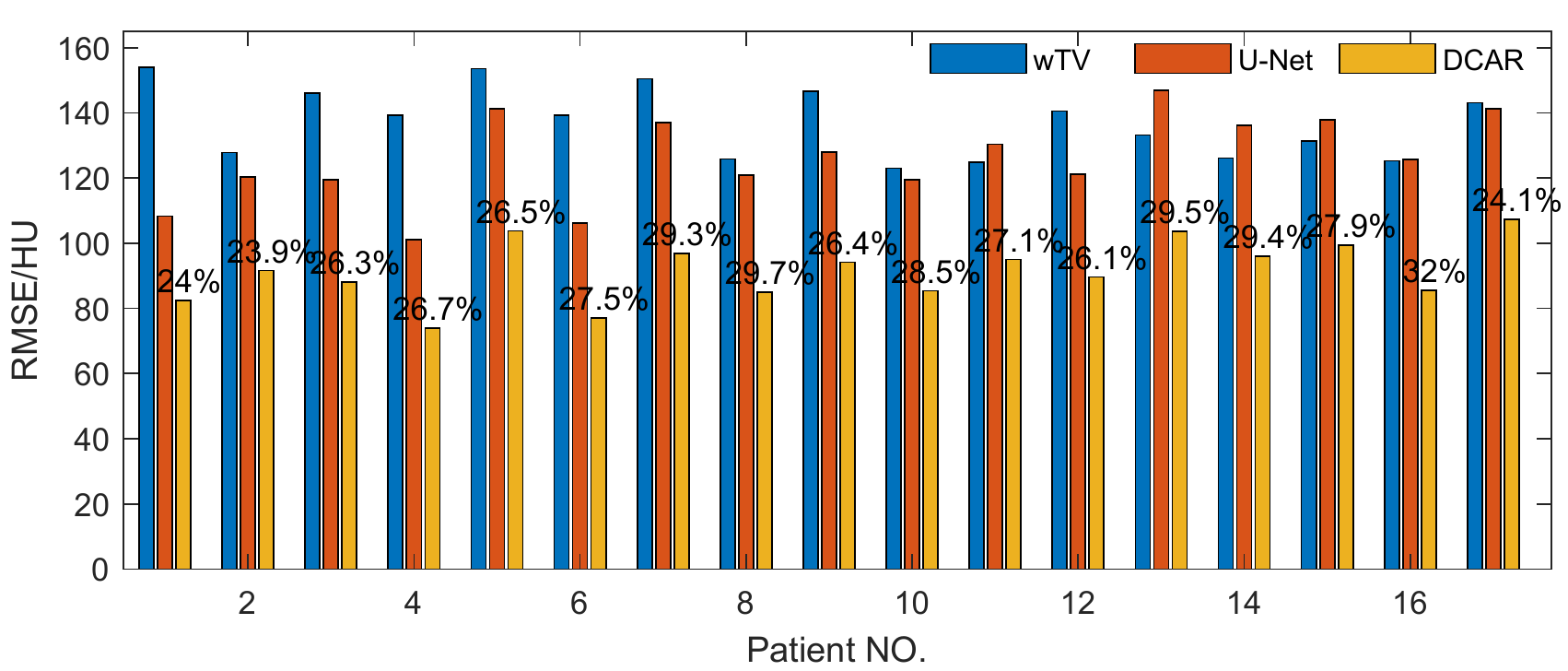}
\end{minipage}
\caption{Comparison of the mean RMSE values by wTV, U-Net, and DCAR for each patient in $120^\circ$ cone-beam limited angle tomography with Poisson noise. The relative improvement of DCAR from the U-Net is marked for each patient.
}
\label{Fig:barPlotMeanValuesPoisson}
\end{figure}

In $120^\circ$ cone-beam limited angle tomography with Poisson noise, the results of three example slices are displayed in Fig.~\ref{Fig:DCARResultsPoisson}. These three slices are from Patient NO.~17, 2, and 8, respectively. In the top row, Fig.~\ref{subfig:reconLimitedP18S128Poi} exhibits a high level of Poisson noise especially for the areas where a lot of X-rays are missing. The Poisson noise is entirely reduced by wTV in Fig.~\ref{subfig:reconwTVP18S128Poi}. However, like the noise-free cases, the top body area is still distorted. Fig.~\ref{subfig:reconUNetP18S128Poi} indicates that the U-Net trained on noisy data is still able to reduce limited angle artifacts. In addition, most Poisson noise is also prominently reduced and only a small portion of it remains. However, many low/median contrast structures, e.\,g. fat and muscles in the area marked by the red arrow, are blurred and cannot be distinguished between each other. Fig.~\ref{subfig:reconDCARP18S128Poi} indicates that DCAR can further reduce Poisson noise and improve low/median contrast structures, as no Poisson noise remains at all and the fat and muscle tissues can be distinguished between each other. The benefit of DCAR is also demonstrated by the RMSE value as it decreases from 138\,HU in Fig.~\ref{subfig:reconUNetP18S128Poi} to 102\,HU in Fig.~\ref{subfig:reconDCARP18S128Poi}. For the slice in the middle row, the U-Net also reduces most of the artifacts and Poisson noise, comparing Fig.~\ref{subfig:reconUNetP2S94Poi} with Fig.~\ref{subfig:reconLimitedP2S94Poi}. However, the cavities in the marked green box in Fig.~\ref{subfig:referenceP2S94Poi} are missing in Fig.~\ref{subfig:reconUNetP2S94Poi}. They are smoothed out by the U-Net. Instead, DCAR is still able to reconstruct most of these cavities, as displayed in Fig.~\ref{subfig:reconDCARP2S94Poi}. For the slice in the bottom row, many dark dots occur in the U-Net reconstruction in Fig.~\ref{subfig:reconUNetP9S140Poi}, due to severe Poisson noise in the limited angle reconstruction in Fig.~\ref{subfig:reconLimitedP9S140Poi}. However, these dark dots are eliminated by DCAR in Fig.~\ref{subfig:reconDCARP9S140Poi}. Except for these example slices, the comparison of the mean RMSE values for wTV, U-Net, and DCAR is displayed in Fig.~\ref{Fig:barPlotMeanValuesPoisson}. The mean RMSE values for wTV stay similar for both the noise-free and noisy cases. However, DCAR achieves more than $24\%$ improvement compared with the U-Net in the noisy case. These remarkable results have demonstrated the robustness of DCAR to Poisson noise in $120^\circ$ cone-beam limited angle tomography.

\section{Discussion And Conclusion}
In the cross-validation experiments, for each test, 16 patients' CT data are used to train the U-Net. Since only 13 slices are chosen from each patient, 400 slices in total are used for training, which is very likely insufficient. Therefore, the U-Net training on such data has a limited generalization ability to test data. That is one potential cause to the dark holes in the U-Net reconstructions in Fig.~\ref{Fig:DCARResults} in the noise-free case. The occurrence of such dark holes make deep learning reconstructions not consistent to measured projection data. DCAR has the ability to improve such reconstructions by constraining them consistent to measured data.

In the noisy case, due to the curse of high dimensional space, noise will accumulate at each layer of the U-Net. Therefore, even if noise has a small magnitude, it still has a severe impact on the output images. That is why the U-Net is not robust to Poisson noise \cite{huang2018some}. In this work, the U-Net is trained on data with Poisson noise. This endows the U-Net to deal with Poisson noise to a certain degree. Fig.~\ref{Fig:DCARResultsPoisson} indicates that the U-Net is able to reduce a certain level of Poisson noise in a manner of smoothing structures. In such a manner, some fine structures are also smoothed out, e.\,g., the small cavities in Fig.~\ref{subfig:referenceP2S94Poi}. In addition, in our experimental setup for the noisy case, the initial photon number without attenuation is relatively low. Hence, the Poisson noise in the FBP reconstruction images is well observed. In some cases, e.\,g. in Fig.~\ref{subfig:reconLimitedP2S94Poi}, the Poisson noise is so strong that the U-Net is not able to reduce it. However, DCAR adapts the SART algorithm using soft-thresholding operators, which is noise tolerant. In addition, the wTV regularization further reduces the influence of Poisson noise as such high frequency noise pattern contradicts a gradient-sparse image, which wTV seeks.

In conclusion, the proposed DCAR method has better generalization ability to unseen data and is more robust to Poisson noise than the U-Net. This is demonstrated by our experiments, achieving significant image quality improvement. Compared to the U-Net, our method reduces the RMSE by more than $10\%$ in the noise-free case and $24\%$ in the noisy case for $120^\circ$ cone-beam limited angle tomography.

\ 

\textbf{Disclaimer:} The concepts and information presented in this paper are based on research and are not commercially available.
%
%

%

\end{document}